\documentstyle[prl,aps,psfig,twocolumn]{revtex}
\newlength{\figwidth}
\def\d{\partial}
\def\o{\omega}

\def\real{{\rm I\kern-.2em R}}
\def\complex{\kern.1em{\raise.47ex\hbox{
            $\scriptscriptstyle |$}}\kern-.40em{\rm C}}
\title{The Longuet-Higgins Phase and Charge Transport in Molecular
Rings}
\author{J.~E.~Avron and J. Berger}
\address{Department of Physics, Technion, 32000 Haifa, Israel}
\begin{document}
\flushbottom
\twocolumn[
\hsize\textwidth\columnwidth\hsize\csname @twocolumnfalse\endcsname
\maketitle
{\begin{abstract}
The Longuet-Higgins-Berry's phase has  remarkable consequences for charge
transport in molecular rings. For
generic (conical) crossing, where the phase is $\pi$, a
vanishing cause can lead to a diverging response in the amount of charge
transport. Away from level crossings, when the phase is $0$, a vanishing cause
leads to a vanishing response. The divergence of the
response  near crossing is  related to, but distinct from,
the divergence that occurs in the generalized susceptibility.   We illustrate
this behavior for quantum models of
molecular rings driven by a running wave of small amplitude at  zero and
finite temperatures.
\end{abstract}  }
{ }]
Consider  a molecular ring, such as a benzene or a
triangular molecule
$X_3$, such as  Na$_3$.  In the Born-Oppenheimer limit of heavy nuclei one can
consider a cycle of deformations where each nucleus is displaced only
slightly
from its initial position and eventually  returns to it. The question that
we want to focus on is: What is the electronic charge transported around the
molecular ring in one such cycle? As we shall explain below, there are  two
cases: If the cycle of atomic deformations can be shrunk to zero without
trapping a point of level crossing (of the electronic energy levels) then one
gets normal behavior in the sense that the weaker  the deformation, the
less the
charge  transported in one cycle. If, however, the cycle of atomic deformations
pinches a point of level crossing, then the smaller the cycle, the {\em larger}
the transported charge. We dub such  anomalous behavior,
where  the weaker the cause the larger the effect, {\em  homeopathic}. We shall
illustrate this for H\"uckel (tight binding) models of molecular rings at zero
temperature.

Homeopathic charge transport is  intimately related to the
Longuet-Higgins \cite{LH} and Berry's phase \cite{berry}: For
time reversal invariant Hamiltonians level crossing is reflected in the
sign of
the electronic wave function undergoing a cycle of deformation.
This  sign is $-1$ if the cycle pinches a (generic) crossing and $1$ if it does
not. The  Longuet-Higgins phase has important consequences for (molecular)
rovibronic spectra in Born-Oppenheimer and Jahn-Teller theory
\cite{mead,mt} and plays a role in molecular
dynamics \cite{cina}. The  theory of adiabatic transport adds the observation
that Longuet-Higgins phase  has  direct consequences also for {\em electronic}
properties, and not only to molecular properties. In particular, a phase
$-1$ implies homeopathic charge transport. There is both theoretical and
experimental evidence that molecular trimers such as Na$_3$
\cite{delcatraz} and
{\it sym}-triazine
\cite{st}  as well as other trimers and  systems have $-1$ Longuet-Higgins
phase.

Adiabatic charge transport near gap closures for {\em infinite chains} has been
studied in  \cite{ressor}.  For these there is no homeopathic divergence.
Instead, a vanishing cause can lead to a {\em finite and quantized} charge
transport when the Longuet-Higgins phase is $-1$. The charge transport in
finite molecular rings  is, therefore, more singular than that in infinite
chains.


The homeopathic behavior that occurs for (out of equilibrium) charge transport
is related to, but distinct from, the divergences that can occur in
thermodynamic equilibrium of generalized susceptibilities at  $T=0$. A
necessary
condition for either, at least for the simple model systems we consider below,
is that  quantum energy levels cross. But, while  thermodynamic
susceptibilities
probe the singularity of the {\em energy surface}
near crossing, non-equilibrium adiabatic transport probes the singularity
of the
{\em surface of eigenstates} near crossings. It is possible for one of these
surfaces to be singular without the other being singular. We shall return
to this
issue below.

As we shall see, finite temperature  introduces a
cutoff of the homeopathic divergence. In some  cases,
a Jahn-Teller instability can  censor the homeopathic divergence even at
$T=0$.

Consider, for simplicity, the  H\"uckel (tight-binding)
model   Hamiltonian  for non interacting electrons in a general triangular
molecule of three identical atoms. %
Although the example of a molecular trimer is special, it turns out that it
describes the generic situation near level crossing.
 The Hamiltonian is the $3\times 3$
Hermitian matrix
\begin{equation}H(a,b,c,\phi)=E_0\,\pmatrix{0&a&\bar \xi c\cr
           a &0&b\cr
           \xi c&b&0\cr},\label{triangle}
\end{equation}
 where $E_0$ fixes the energy scale and  $a,b,c$ are (dimensionless, real)
hopping
amplitudes associated to the three  bonds  of the triangle. (The triangle
is not
necessarily equilateral.) We assume that $a,b,c$ are all positive. We can, and
shall, use $a,b,$ and $c$ as {\em local} coordinates in the space of (internal)
configurations of a trimer. $a,b,$ and $c$ are actually {\em not} good
coordinates globally.
For good global coordinates see
\cite{mead}. However,  global subtleties need not concern us here since we
consider only small deformations.

 $\xi=\exp i\phi$, with $\phi$ an
auxiliary phase variable  associated with a {\em fictitious } Aharonov-Bohm
flux tube which carries flux
$\phi$ and threads the molecule. The explicit form of Eq.~(\ref{triangle})
involves a choice of gauge for the flux tube. (We shall  consider
observables that are independent of this choice.) The role of
$\phi$ will become clear below. Deformations of the molecule change the hopping
amplitudes, and a closed cycle of deformations is a  closed path in the three
dimensional space  whose points are the hopping amplitudes
$(a,b,c)\in\real_+^3$.  Such a closed path is shown in Fig.~1. For notational
convenience we denote by $X$ the triplet $(a,b,c)$.

The reason for introducing the fictitious flux tube $\phi$ is to define the
current operator which circulates in such a ring. For the choice of gauge
we have
made for the Aharonov-Bohm flux, the current is associated with a single bond,
the
$c$ bond, and is
\begin{equation} (\d_\phi H)(X,\phi)=c E_0\pmatrix{0&0&-i\,\bar \xi \cr
           0 &0&0\cr
           i\,\xi &0&0\cr}.
\label{curr_tri}
\end{equation} This is the sole role of $\phi$ and in all our calculations we
shall eventually set $\phi=0$, which is the case with no flux at all. In this
case the Hamiltonian
$H(X, 0)$ is real and therefore time reversal invariant.
The observable associated to the circulating current,
Eq.~(\ref{curr_tri}),  is pure imaginary when $\phi=0$ and so odd under
time reversal. Because the model for $\phi=0$ is time reversal invariant,
there are no
diamagnetic (persistent)  currents in any eigenstate. Let  $P(X,\phi)$ be  a
spectral projection for
$H(X,\phi)$, i.e. $H(X,\phi) P(X,\phi)= E(X,\phi)P(X,\phi)$ with
$E(X,\phi)\in\real$ an eigenvalue.  The vanishing of the persistent
currents is the statement
$Tr\, (P\partial_\phi H)\vert_{\phi=0}=0$ for all $X$.

We are interested in the current that flows around the molecule when it is
slowly driven so that $X$ traces a closed path in parameter space, as
e.g. in Fig.~1. In the theory of adiabatic transport this current is related to
the adiabatic curvature. The basic equation for the
expectation value of the current at time $t$ and zero flux, reads
\cite{tknn,avron,besb,kohmoto,kunz,niu,seiler,thouless,stone,resta}:
\begin{equation}
 Tr\,( P_t\,\partial_\phi H)=
 Tr\,\Big(\Omega_{\phi X}(P) \Big)\cdot\dot X
+O(1/\tau^2).\label{basic}
\end{equation}
$X$ stands for the triplet $a,b$ and $c$
and the dot denotes  time derivative.
$P_t$ is a solution of the quantum evolution equation, with the (adiabatic)
time dependent Hamiltonian $H(X,\phi)$ and with initial condition that
$P_{t=0}$ is an eigenstate.  $P(X,\phi)$ is an instantaneous spectral
projection
for the instantaneous $H(X,\phi)$.
$\Omega_{\phi X }(P)=-i\, P\,[\partial_\phi P,\partial_X P]\,P$ is the $\phi X$
component of the adiabatic curvature \cite{berry}. $\tau $ is the
time scale so that the adiabatic limit is $\tau \rightarrow
\infty$.
The charge transported around the ring in one cycle, in the adiabatic limit, is
\begin{equation}
Q=\int_0^\tau dt\,  Tr\,( P_t\,\partial_\phi H)=
\oint Tr\,\Big(\Omega_{\phi X}(P) \Big)\,d X.\label{Q}
\end{equation}
Analysis of the characteristic equation of Eq.~(\ref{triangle}) shows that
level
crossing can occur  only if $a=b=c$ and $\xi=\pm 1$. For $\xi=1$, the case we
consider here, the simple  eigenvalue is
$2aE_0$ (the top state if $E_0$ is positive) and the
corresponding eigenvector is
$|0\rangle= \frac{1}{\sqrt 3}\,(1,1,1)$. The two fold degenerate eigenvalue
is $-aE_0$ and the projection to its subspace is $1- |0\rangle \langle
0|$. For
$\phi=0$, level crossings occur on a ray in $X$ space.

The essence of homeopathic behavior is the following. Suppose first that the
cycle of deformation does not pinch level crossing. (For the top state this
holds for any closed cycle in the positive quadrant.)  The  adiabatic curvature
is smooth and bounded along the path and the right hand side of Eq.~(\ref{Q})
is of the order of the {\em area} of the cycle of deformations (by Stokes
formula). $Q$ goes to
zero when
the cycle shrinks to zero. Now suppose that the cycle of deformation pinches
the ray of level crossing, e.g.
\begin{eqnarray} a(x)&=& 1 + \o \bar x + \bar\o x; \ \ b(x)= 1 + \bar x
+  x;\nonumber \\
c(x)&=& 1 +\overline{\o x} + \o x,
\label{deformation}\end{eqnarray}
where $\o$ is here a (complex) cube root of unity and
 $x\in\complex$ runs on a small circle in the complex
plane surrounding the origin, see Fig.~1. As the cycle shrinks,
the adiabatic curvature in Eq.~(\ref{Q}) diverges quadratically,
like a monopole \cite{berry,simon}, while the length  of
the cycle shrinks only linearly. It follows that now $Q$ {\em diverges} as
the cycle shrinks to zero, i.e. $Q=O\left(\frac{1}{|x|}\right)$.

In the simple case at hand the leading divergence characterizing
homeopathic
behavior can be calculated explicitly. The degeneracy splits in first order of
perturbation theory, both in $\phi$ and in $x$, and the local behavior near
crossing of the two bottom states is given by  the $2\times 2$ matrix
\begin{equation}
\frac{E_0}{3}\,\pmatrix{-\sqrt{3} \phi &6\o \bar x\cr 6\bar \o  x&\sqrt{3}
\phi}.\label{tbt}
\end{equation}
This matrix has the form of a Berry spin 1/2 model and the adiabatic
curvature of its two states is explicitly computable. So, to leading order, the
equation for  the charge $Q$, Eq.~(\ref{Q}), reduces to \begin{equation} Q=\pm
\frac{\sqrt{3}}{12} \,\oint\frac{1}{|x|}\
d \ Angle,\label{angle}
\end{equation}
where ${Angle}$ is the angle swept by
$x$
as it moves around the origin in the complex plane. The $\pm$ signs refer
to the
ground and first state respectively. A  simple formula is obtained for a
circular orbit, $|x|=const$, where
$Q=\frac{\sqrt{3} \pi}{6|x|}$. Evidently, the smaller the cycle that
pinches the
degeneracy,  the more charge it transports and $Q\to\pm\infty$ as $|x|\to 0$.

It may be worthwile to explain which aspects of Eq.~(\ref{angle}) are general
and hold for any generic two level crossing and what is special for the
explicit model we consider. The overall constant $\sqrt{3}/12$ is
special for the model. What is general is that the divergence scales like an
inverse power of the distance from crossing. This can be seen be noting that
for any (generic) two level crossing the adiabatic curvature diverges like the
field of a monopole, i.e. like $|x|^{-2}$, and therefore a line integral
on a
loop of length $O(|x|)$ will scale like the potential of a monopole.

It should be stressed that this result {\em does not} imply that the current is
large. In Eq.~(\ref{basic}) only the ratio of the current to the rate of
driving
is large. The current is not large because,  as the circle is shrunk, the
rate of
driving must also decrease in order for the adiabatic theory to apply.

It may be worthwhile to point out that the charge
$Q$ in Eq.~(\ref{angle}) even though a geometric, {\em is not}
a Berry's phase (being a line integral, rather than a
surface integral, of the curvature).

We now return to the question of how is homeopathic behavior of transport
related to
the divergence of the susceptibility near level crossing. At $T=0$
the generalized susceptibility  matrix $\chi$ of a family of Hamiltonians,
$H(X,\phi)$, that depend parametrically on $X$ and $\phi$ is the {\em
symmetric} matrix  of second derivatives:
$$
\chi_{XX}(X,\phi)= \frac{\partial E}{\partial x_i\partial x_j}(X,\phi),\
\chi_{X\phi}(X,\phi)= \frac{\partial E}{\partial x_i\partial \phi}(X,\phi)
,
$$
where  $E(X,\phi)$ is the ground state of $H(X,\phi)$. $\chi$ is a
thermodynamic (equilibrium) property and as such it
depends {\em only} on the the energy surface $E(X,\phi)$. The adiabatic
curvature  is associated with an {\em anti-symmetric} matrix whose
components are
\begin{eqnarray}
Tr\Omega_{XX}(X,\phi)&=& -i\,Tr(P[\partial_{x_i} P,\partial_{x_j}
P]),\nonumber \\
Tr\Omega_{X\phi}(X,\phi)&=& -i\,Tr(P[\partial_{x_i} P,\partial_{\phi} P]).
\end{eqnarray}
It describes transport coefficients
which {\em can not} be determined from the the ground state energy (at $T=0$)
or a thermodynamic potential (for $T>0$).  This is evident from the
formula for the adiabatic curvature, which is determined by the projection
$P(X,\phi)$ on the ground state and not on its energy $E(X,\phi)$.

Eigenenergies and eigenstates are, of course, related. Away from
level crossings, both $E$ and $P$ are smooth functions of the parameters.
However, besides that, the susceptibility and the curvature are
essentially
independent quantities. For example, from  Eq.\,(\ref{tbt})  the energy surface
of the ground state is
$E(\phi,x)= -E_0\,(1+\sqrt{\phi^2/3 +4 |x|^2})$. So, while
$\Omega_{X,\phi}(X,\phi=0)$
diverges at the crossing, the corresponding component of the
susceptibility,
$\chi_{X,\phi}(X,\phi=0)$  {\em vanishes identically} and, while
$\chi_{\phi\phi}$ diverges at crossing, the corresponding curvature
$\Omega_{\phi\phi}$ {\em vanishes identically}.

Finite temperature provides a cutoff to the homeopathic divergence. Two
nearly
crossing states, which are at distance $\varepsilon$ apart in energy,
transport
opposite charges $Q=\pm O(1/\varepsilon )$.  At finite temperature $T$
the two nearly crossing states will be nearly equally populated with a bias of
$O(\varepsilon/T)$ towards the lower state.
For the triangular molecule undergoing a cycle of deformations,
the leading behavior of the charge transport at low temperature is:
$$Q(T)=\frac{\pi E_0}{\protect{\sqrt{3}}k_B T},$$where $k_B$ is
Boltzmann's constant.
The constants in this formula are special for the model. In general,
for generic two level crossings, one can conclude that the total charge
transport will  approach {\em a finite limit} as one approaches the
crossing so that:
\begin{equation}
Q(T)=O(1/T).
\label{curie}
\end{equation}
 The $1/T$ law is reminiscent of Curie's law.

Similar analysis can be made for a
necklace of $p$ equivalent atoms with $p$ an arbitrary integer larger
than 2. (For benzene $p=6$.) The
tight  binding Hamiltonian is a $p \times p$ Hermitian matrix with nearest
neighbors hopping only. Now suppose that such a ring is deformed by means of a
running sinusoidal wave of commensurate wavelength, i.e.  the hopping amplitude
between the $k$-th and
$(k+1)$-th atom is time dependent and is
given by $1+2\kappa\cos[ 2\pi(jk/p-t/\tau)]$, where $\kappa$ (which is
held fixed) stands for the amplitude
of the  distortion and $j$ is a natural number
smaller than $p$. (This is an analog of Eq.~(\ref{deformation})).
When  $\phi=\kappa=0$, there are twofold degeneracies at energies $2\cos
(2m\pi/p)$ for $m=\pm 1, \dots, \pm [p/2]$.  The new feature of this model is
that the order of perturbation theory that  splits the degeneracies depends
on
$|m|$, $p$ and $j$, and can be large if $p$ is large. As a consequence, the
singularity of the adiabatic curvature near crossing can be quite strong
(without the susceptibilities being singular).
The charge transported can be calculated here as well, but the details  will be
described elsewhere. The result is that the charge transported in the pair of
nearly crossing states near energy $2\cos (2m\pi/p)$ is:
\begin{equation}
Q=\pm \frac{\pi q \sin m\theta}{p\kappa^q \cos(m-j/2)\theta}
\left( \prod_{k=1}^{q-1} \frac{\cos(m-jk)\theta-\cos m \theta}
{ \cos(m+j/2-kj)\theta}\right).
\label{QQ}\end{equation}
$q\ge 1$ is  the order of perturbation theory that splits the degeneracy.
$q$ is the smallest natural number such that $qj$ mod $p$ equals either
$2|m|$ or $p-2|m|$. In the first case $m=|m|$ in Eq.~(\ref{QQ}) and, in
the second case, $m=-|m|$.
The numerator and the denominator never vanish under the conditions that
lead to this equation.  $\theta$ is
shorthand for $2\pi/p$. We see that the amplitude of the perturbation $\kappa$
enters this expression  with a negative exponent: this is the homeopathic
effect. The sign of $Q$ reflects the fact that each member of the
pair of nearly crossing states transports charge in opposite sense.
Under a complete cycle $t \rightarrow t+\tau$ the electronic eigenstate
acquires a phase factor $(-1)^q$.
Eq.~(\ref{curie}) is still obeyed.

The molecular models discussed so far are prototype models of {\em finite}
quantum systems.  It is natural to inquire what, if any, of the homeopathic
behavior survives for infinite, macroscopic systems. A prototype of such a
system is the infinite one dimensional  chain with finite
electron density. When the Fermi energy lies in a gap such a chain is
nominally an insulator and the theory of adiabatic transport applies. For non
interacting electrons and a periodic chain, the transport behavior can be
analyzed using standard, single electron, techniques.   It turns out
that there is no homeopathic divergence of transport. What survives of the
homepathic behavior is that an arbitrarily small cycle of deformation can lead
to a finite (nonzero) and quantized charge transport. This requires that
the deformation pinches a gap closure. The quantization of transport (at
$T=0$) comes as  it does in charge pumps
\cite{niu,thouless} and the Hall effect \cite{tknn}.  The theoretical framework
presented above sheds light on the numerical findings of
\cite{ressor} who found divergence of the transport coefficients in Hubbard
models of Perovskite chains.

Until now we have only considered the electronic part of the Born-Oppenheimer
theory. The inclusion of the rovibronic part is, in general, a formidable
problem even for a molecular trimer where the intricacies of
the three body problem come into play. We refer to \cite{mead} for what is
known in general and to \cite{cina,obrien,ham,jahnteller}, for models.  We shall
content ourselves here with the classical limit for the nuclear dynamics.

If the ground state of the undeformed molecule is degenerate, the total energy
may be decreased by deforming the molecule. This is the classical Jahn Teller
\cite{jahnteller} instability. This is the case if the  elastic energy is
proportional to the square of the amplitude of the  deformation (i.e. is
harmonic) and if the degeneracy lifts to first order in the deformation. If the
Jahn Teller energy functional has a unique minimizer which breaks the
degeneracy, then a sufficiently small cycle of deformations around the ground
state will not encircle also  the point of  level crossings. In this case the
homeopathic behavior (at $T=0$) is censored at the ground state by the Jahn
Teller instability and there is no divergence of the charge transport in the
limit of an infinitesimal cycle of deformation.

\section*{Acknowledgments}
This work was partially supported by a grant from the Israel Academy of
Sciences, the Deutsche Forschungsgemeinschaft, and by the Fund for Promotion of
Research at the Technion.  J. B. was also supported by the US-Israel Binational
Science Foundation. We have benefited from conversations with R. Englman.

\begin{figure}[thb]
\centerline{\psfig{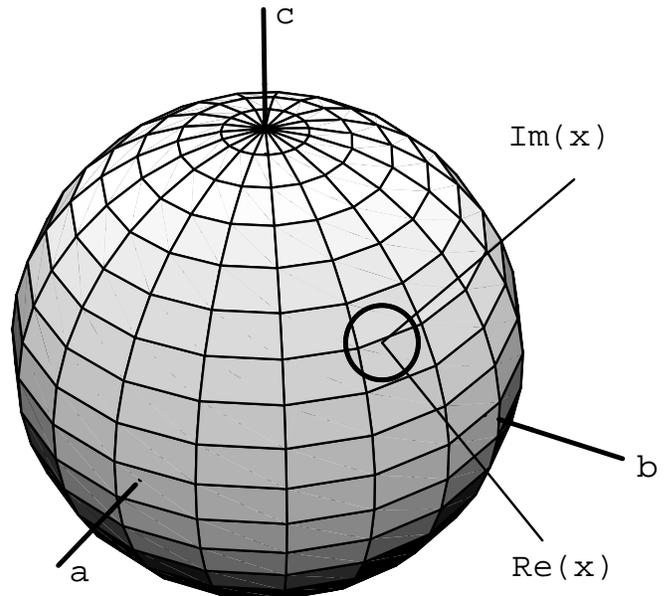}}
\caption{A cycle of deformation for a molecule with three atoms.}
\end{figure}
\end{document}